# Effect of strain on phase separation and de-vitrification of magnetic glass state in thin films of $La_{5/8-y}Pr_yCa_{3/8}MnO_3$ (y=0.45)

V.G. Sathe, Anju Ahlawat, R. Rawat and P. Chaddah

*UGC-DAE Consortium for Scientific Research,*

*University Campus, Khandwa Road, Indore-452001 India*

***Abstract***

We present our study of effect of substrate induced strain on the $La_{5/8-y}Pr_yCa_{3/8}MnO_3$ (y=0.45), thin films grown on $LaAlO_3$, $NdGaO_3$ and $SrTiO_3$ susbtrates that show large scale phase separation. It is observed that unstrained films grown on $NdGaO_3$ behave quite similar to bulk material but the strained film grown on $SrTiO_3$ show melting of insulating phase to metallic phase at low temperatures. However, the large scale phase separation and metastable glass-like state is observed in all the films despite difference in substrate induced strain. The measurements of resistivity as a function of temperature under cooling and heating in unequal field (CHUF) protocol elucidates the presence of glass-like meta-stable phase generated due to kinetic arrest of first order transformation in all the films. Like structural glasses, these magnetic glass-like phase shows evidence of devitrification of the arrested charge order-antiferromagnetic insulator (CO-AFI) phase to equilibrium ferromagnetic metallic (FMM) phase with isothermal increase of magnetic field and/or iso-field warming. These measurements also clearly brought out the equilibrium ground state of this system to be FMM and metastable glass-like phase to be AFI phase.

**Introduction:**

The current leading efforts to understand colossal magneto resistance (CMR) effects in manganites centers around inhomogeneities and first order insulator-metal transitions [1]. The first order transition gets significantly broadened with increasing quenched disorder present in the system. $La_{5/8-y}Pr_yCa_{3/8}MnO_3$ (LPCMO) is one such highly studied system, where substitutional disorder leads to phase separation at micrometer length scale. Many researchers believe it to be a result of quenched disorder due to variation in ionic size of the constituent elements [2]. On the other hand magnetic force microscopy studies elucidated fluid like growth of the ferromagnetic metal (FMM) phase [3]. Presence of coexisting phases is an essential phenomena across any first order transition and disorder makes the transition broad making coexistence of different phases possible for a sufficiently broad temperature and spatial range. In LPCMO the two coexisting phases are charge order antiferromagnetic insulator (CO-AFI) and FMM. These series of materials show re-entrant transitions in magnetization and resistivity measurements that are not well understood [4,5,6,7]. Uehara *et al* [4] studied the relaxation behavior and observed that the value of resistivity strongly depends on the cooling rate. They described the phase separated states as strain glass states [5]. Sharma *et al* also described the low temperature state, in which CO-AFI and FMM phase coexist as strain glass like magnetism [5] while Ghivelder et al classified the phase separation as dynamic phase separation and frozen phase separation [8, 9]. Dhakal *et al* [6] studied the magnetic field vs. temperature (H-T) phase diagram and further observed these two distinct types of phase separation, a strain-liquid (dynamic phase separation) and a strain-glass (Frozen phase separation) regions. It is argued that in phase separated scenario the two phases are of nearly same energy and the balance can be changed by external stimuli like magnetic field, laser photons etc. It is shown that CO-AFI state of this system can be converted to FMM state by application of external stimuli like magnetic field, current, voltage, laser photons etc. at low temperature. However, the system does not transform back to its original AFI state when the external field is withdrawn isothermally [7]. This naturally raises the question about the equilibrium state, whether the CO-AFI is equilibrium state or the remnant state FMM is the equilibrium state? Thus a unified understanding is lacking that can explain the different phase separated phases and role of strain on the phase fraction ratio below phase transition temperature apart from describing correct ground state of the system.

It has previously been suggested that strain parameter can tune the metal-insulator phase fraction in this LPCMO series of compounds [10]. Further, recently it is argued that the observed micro(meso) meter scale phase separation owes their existence to extrinsic causes like strain [11]. It is also remarked that 'kinetic arrest' could be another possible reason for the observed micrometer scale phase separation in manganites. Glass like arrest of kinetics across first-order magnetic transition has been reported in materials ranging across intermatallics and colossal magnetoresistance mangnites [12,13,14,15,16,17,18, 19] and is termed as magnetic glass. The first order phase transition can be fully or partially arrested at low temperatures; the arrest occurs as one cools below a temperature $T_g$ called magnetic glass transition temperature that is concurrent to the classical liquid glass transition temperature when one cools the liquid at a higher rate so that one passes the transition temperature without crystallization. And this frozen state gets "de-arrested" over a range of temperature as one warms showing devitrification [13, 16, 17]. Depending on the system, the glass like arrested state can be antiferromagnetic [13, 14, 15, 16] or ferromagnetic [12, 13, 17, 18]. Thus it is interesting to study the role of strain and 'kinetic arrest' in LPCMO compound that shows micrometer scale phase separation. One way of introduction of strain in the compounds is to grow highly oriented thin films on different substrates. In the present study LPCMO films are grown on three substrates of $NdGaO_3$ (NGO), $SrTiO_3$ (STO) and $LaAlO_3$ (LAO) that provide three different values of strain in the films.

In this paper, it is shown that in this system the CO-AFI phase obtained as a majority state on cooling in H=0, shows all the characteristics of a glassy state including devitrification on warming. The recently designed 'cooling and heating in unequal fields' (CHUF) protocol [20] has been used to show that devitrification occurs whenever the sample is warmed in a magnetic field ($H_W$) that is higher than the field ($H_C$) it was cooled in; a reentrant CO-AFI to FMM to CO-AFI transition is seen on heating as devitrification at $T_g$ is followed by the first order transition at T** analogous to melting at $T_m$ [20]. Here we show that while T** rises with increasing H (as expected), $T_g$ falls with rising H similar to that observed in $Pr_{0.5}Ca_{0.5}Mn_{0.975}Al_{0.02}5O_3$ (PCMAO) system [21].

## Experimental:

In this study, we have used oriented thin films of $La_{5/8-y}Pr_yCa_{3/8}MnO_3$ (y=0.45) grown on commercially procured $SrTiO_3$, $NdGaO_3$, and $LaAlO_3$ single crystal substrates using Pulsed laser deposition system. We have chosen this well-studied y=0.45 composition which shows a photo induced transition [7]. The lattice parameter of the bulk material nearly matches with the NGO (110) substrate (3.835 Å) thus the strain imparted on the film is very low (<0.4%). On the other hand the lattice parameter of the LAO (100) substrate (3.79 Å) is smaller than the bulk lattice imparting in-plane compressive strain of 1.4% and the lattice parameter of the STO (100) substrate (3.905 Å) is larger than the bulk giving in-plane tensile strain of 1.6%. The polycrystalline sample for making target in the pulsed laser deposition process was prepared by standard solid state reaction method and well characterized by Rietveld analysis of the x-ray powder diffraction data and Raman spectroscopy. The lattice parameters obtained from the Rietveld analysis (orthorhombic unit cell) are found to be a=5.435 Å, b=7.663 Å and c=5.4285 Å. The films were prepared using 248 nm KrF laser source operating at 3 Hz frequency and ~2 J/cm$^2$ energy density on the target. The thickness of all the films was kept around 200 nm. The chamber was evacuated initially to 1x10$^{-7}$ Torr pressure while depositions were carried out in the partial oxygen pressure of 300 mTorr and the substrate was kept at 700°C. After deposition the films were annealed in the same oxygen pressure for 30 minutes before slow cooling to room temperature. The films were well characterized by x-ray diffraction method. The resistivity (ρ) measurements were carried out by standard four probe method where magnetic field and current directions were kept parallel to each other. The typical current value was 1 μA for all the samples.

## Result and Discussions:

Figure 1 shows standard θ-2θ x-ray diffraction patterns of films deposited on STO, NGO and LAO respectively. The diffraction data shows that the films are perfectly oriented and of a single chemical phase and very good crystalline quality. A small peak is observed around 42.5 degrees that corresponds to contribution to the main peak from Cu-k$_\beta$ line. From peak positions of substrate and films the strain parameter is calculated. From the position of x-ray diffraction peaks of films and substrate it is observed that the film grown on LAO have in-plane compressive strain while film on STO possess in-plane tensile strain. The film grown on NGO has very small amount of strain and the lattice parameter nearly matches with that of bulk compound.

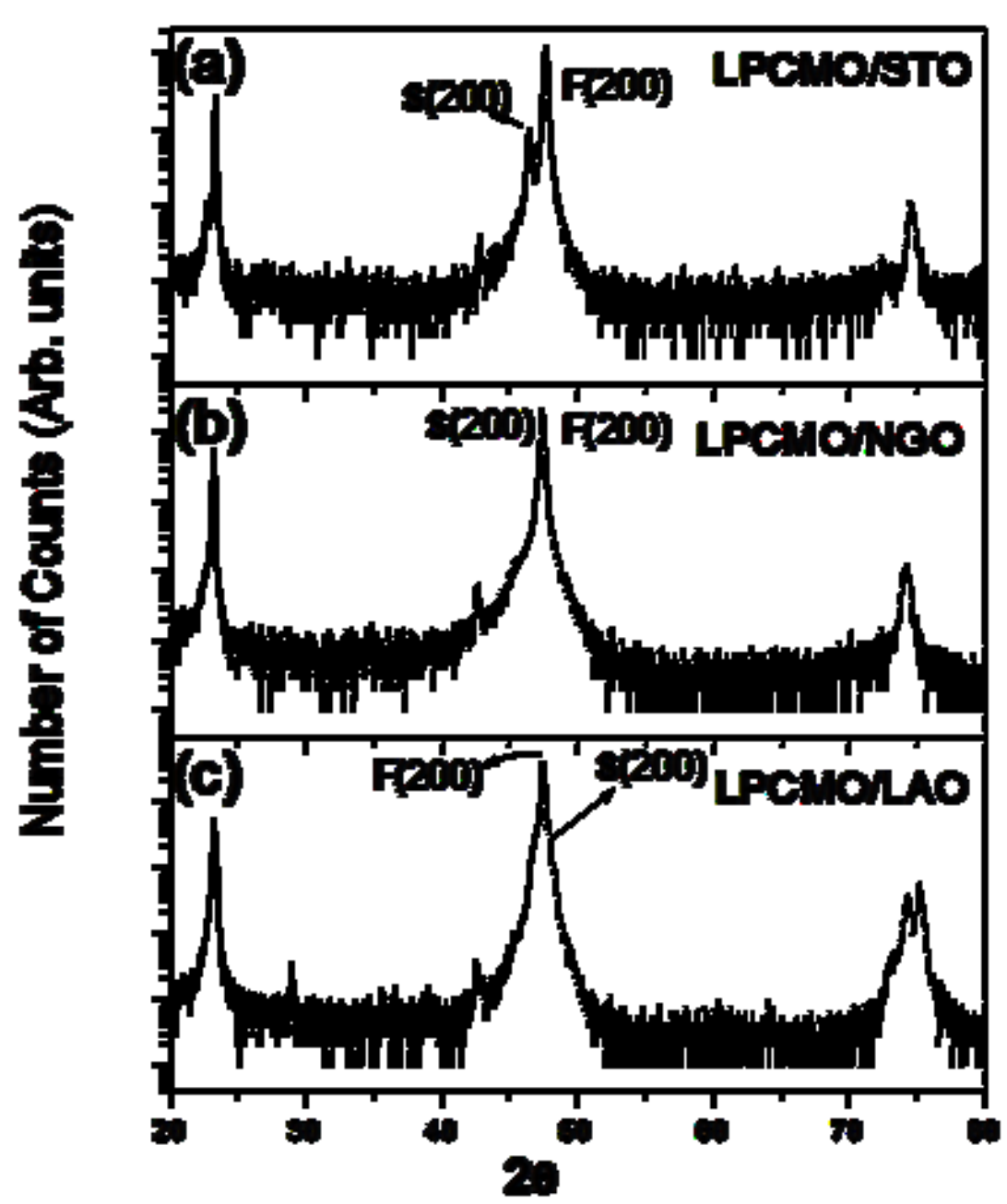

**Figure 1:** X-ray diffraction pattern of thin films of LPCMO grown on (a) STO, (b) NGO and (c) LAO substrates respectively (top to bottom). The data is collected in normal powder diffraction geometry. S(200) and F(200) denotes (200) peak of substrate and film respectively.

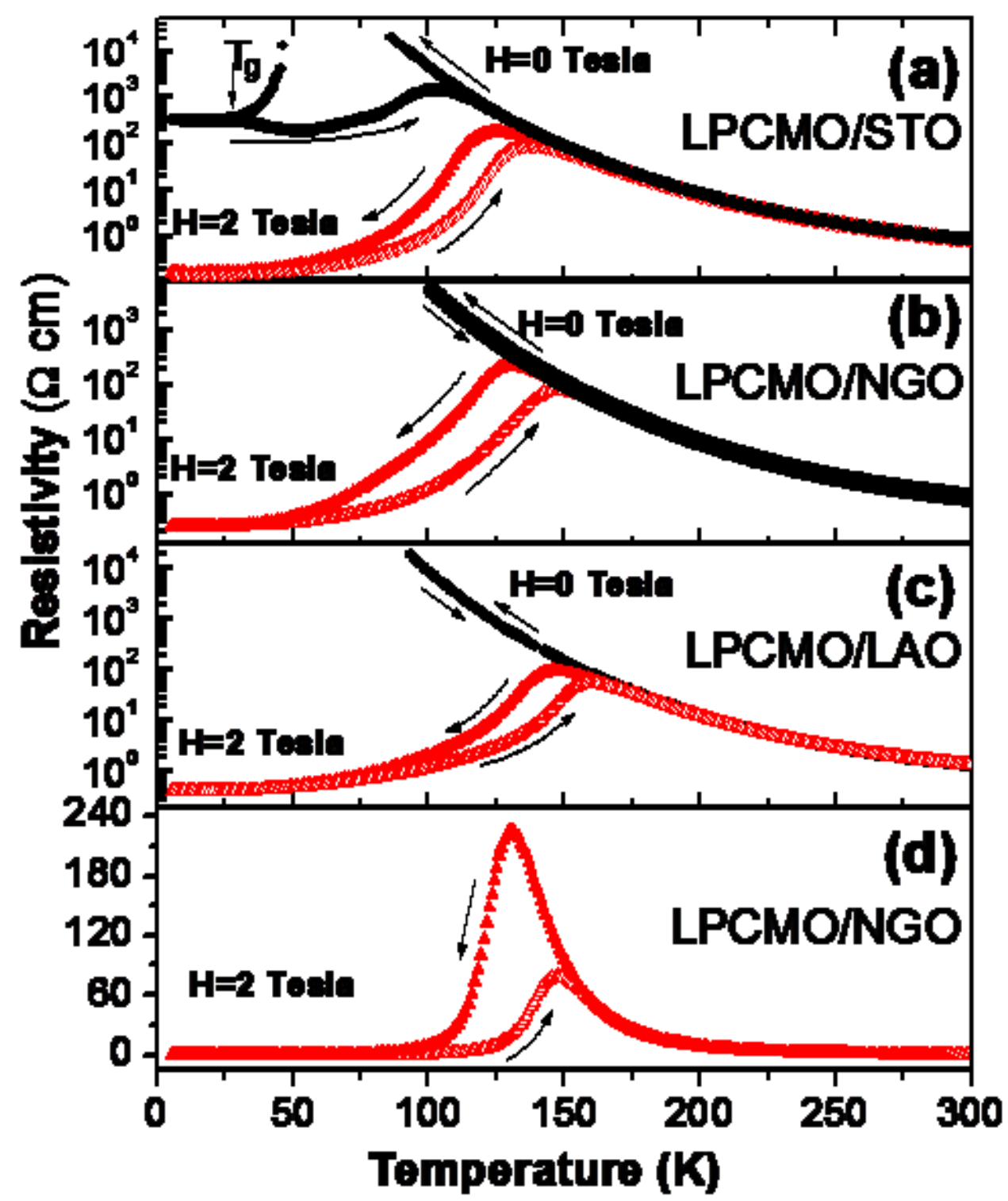

**Figure 2:** Resistivity vs temperature curves for thin films of $La_{5/8-y}Pr_yCa_{3/8}MnO_3$ (y=0.45) on (a) STO substrate (b) NGO substrate and (c) LAO substrate for H=0 T and 2 T cooling and subsequent warming (d) NGO substrate for 2 T Cooling and subsequent warming in linear scale. Cooling and warming directions are indicated by arrows.

In order to study the metal-insulator transition and the associated phase separation phenomena, detailed resistivity measurements as a function of temperature are carried out. The CO-AFI state and FMM transformation is reflected in resistivity measurements with orders of magnitude change when CO-AFI (high resistance) state is transformed to the FMM (low resistance) state through first order phase transition.

The ρ vs T data for the three LPCMO films grown on STO, NGO and LAO are shown in figure 2 for cooling in 0 Tesla to 5 K and subsequent warming, and cooling in 2 Tesla to 5 K (FCC) and subsequent warming (FCW).

The effect of substrate induced strain is clearly seen in the resistivity data. For films on NGO and LAO [figure 2 (b) and (c)] the zero field resistivity showed insulating behavior down to 100 K and remains higher than our measurement limit down to 5 K. The warming curve for these samples overlapped with the cooling curve. However, the cooling and warming curve of the resistivity for H = 2 T showed a clear insulator to metal transition and large hysteresis as highlighted for the film on NGO substrate in figure 2d where resistivity is plotted on a linear scale. This indicates first order transition from AFI to FMM phase under applied magnetic field. On the other hand the films on STO showed a first order transition even for H=0. The transition temperature $T_C$ for H=0 during cooling can not be specified precisely as the resistance value is beyond our measurement limits but it lies near 50 K. The zero field measurements also showed large hysteresis between cooling and warming curve confirming first order transition. The transition shifts to higher temperature when 2T field is applied and the transition temperature is closed to that observed for NGO for the same field value. In short, the least strained film (<0.4% lattice mismatch) grown on NGO showed behavior similar to bulk material [22]. The behavior also matches to that reported in ref. [7] on the exactly same composition film grown also on NGO substrate. The in-plane compressed film grown on LAO (1.4% lattice mismatch) showed enhanced CO-AFI phase fraction while the in-plane tensile strained film grown on STO (1.6% lattice mismatch) showed suppression of CO-AFI phase with a metal-insulator transition near 50 K. Thus it is shown that the strain does affect the observed phase coexistence in H=0.

Uehara *et al* [22] have done a detailed study for various y; for cooling in 0T they observe a FMM phase at low T for y<0.275; a partial conversion to FMM for 0.3<y<0.4 and no conversion to FMM for y>0.41. Bulk samples corresponding to our y=0.45 film show no insulator to metal transition [7, 22]. It must be noted, however, that detailed measurements following novel paths in H-T space established that y=0.41 has an FMM ground state in H=0 [14, 15, 23, 24]. With this background we use the recently designed CHUF protocol to investigate what the H=0 ground state is for both our unstrained (NGO) and strained (STO, LAO) thin films.

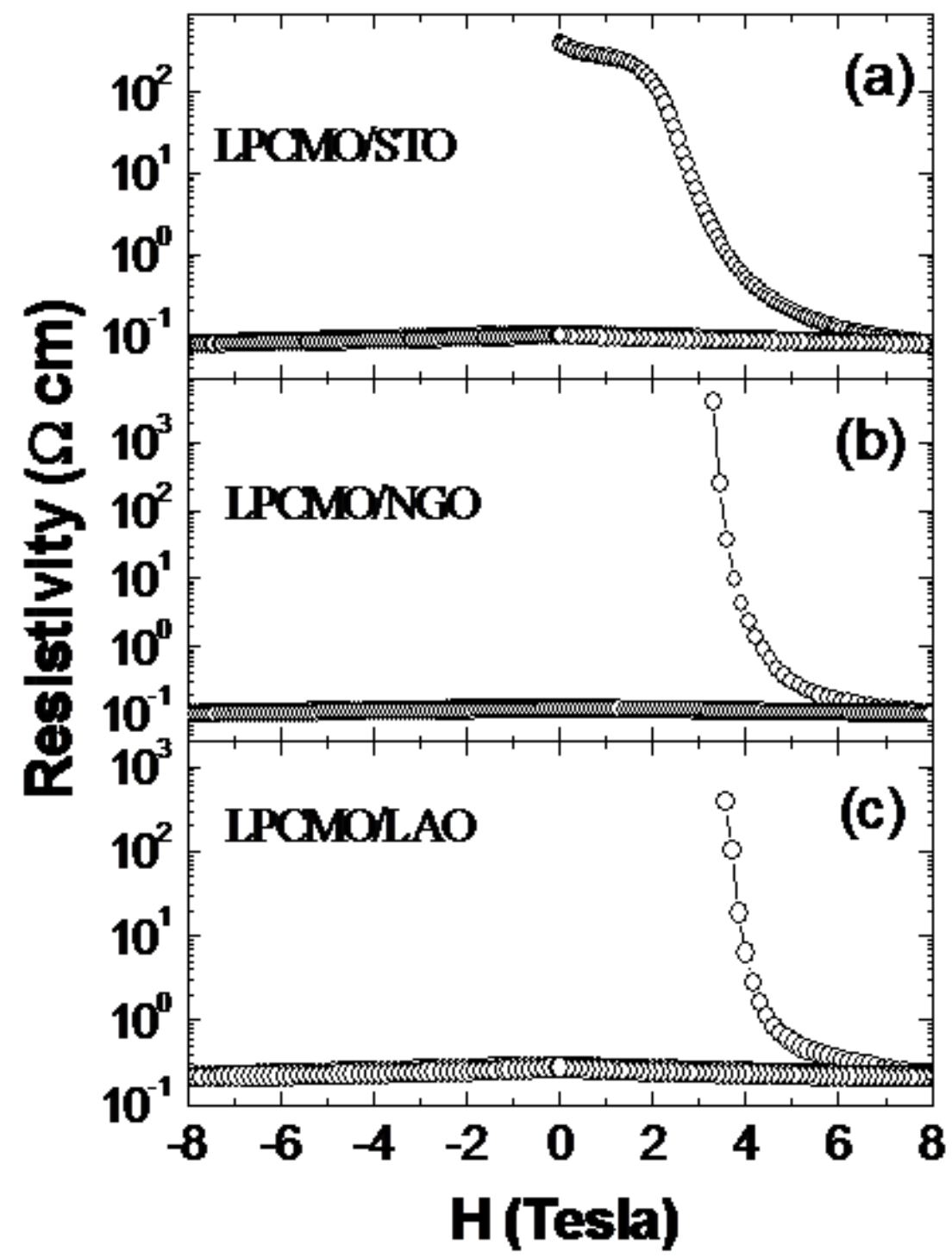

**Figure 3:** Magneto resistance data of (a) LPCMO/STO film, (b) LPCMO/NGO film and (c) LPCMO/LAO film collected at 5 K.

As previously reported [8,22,25, 26, 27, 28], this compound behaves as phase separated system below $T_C$, with coexisting CO-AFI and the FMM phases. The zero field cooled state of LPCMO/STO sample below ~30 K shows an entirely different behavior. The resistivity is nearly constant below this temperature both in cooling and warming curve [see figure 2(a)]. This suggests that the sample is frozen in a metastable state. This temperature is called as freezing temperature $T_g$. Such an anomalous behavior in H=0 resistivity curve below 30 K has been reported previously by many workers and attributed to different mechanisms. Ghivelder *et al* [8] defined this state as a highly blocked state, with a small and almost time-independent fraction of FM phase in a matrix of CO-AFI phase. Sharma et al [5] labeled this state as strained glass state showing highly glassy nature of the charge/spin degrees of freedom and suggested that the reentrant transition could be caused by a small free-energy difference between the competing FM and CO states. Recently, it has been shown in these families of compounds [23, 24] that the static phase coexistence persisting to lower temperature with out any change in phase fraction and is attributed to glass-like arrest of transformation kinetics that is frozen at $T_g$. It is worth mentioning here that this 'magnetic glass' state is completely different from spin-glass, re-entrant spin-glass or cluster glass. In magnetic glass, long range structural and magnetic order is always maintained but the kinetics of first order transition is arrested and the high temperature magnetic phase is retained below $T_g$. This is similar to the conventional glass where the high temperature disordered phase is arrested below glass transition temperature avoiding first order liquid to crystalline phase transition. In real systems, particularly like LPCMO where the doping induced disorder is large, the first order transition is very broad, spurring the partial transformation of the high-T phase upon cooling while the remaining fraction falls out of equilibrium and persists down to lowest temperature as glass like arrested state.

In order to further investigate the nature of phase coexistence and glass like behavior below ~30 K observed in figure 2 detailed magnetoresistance (MR) measurements were carried out by following various paths in H-T space. Figure 3 shows the isothermal magnetoresistance measurement at 5K for all the three samples. For these measurements samples were cooled under zero field to 5 K and magnetic field is applied isothermally from 0 ➔ +8 Tesla (virgin curve) and then +8 ➔ -8 Tesla➔ +8 Tesla (envelope curve). As can be seen from this figure, where resistivity is plotted on log scale, resistivity of films on NGO and LAO becomes measurable for magnetic field higher than ~3 Tesla and decreases by 4 order of magnitude by further increase in magnetic field up to 8 T. This indicates a magnetic field induced transition from CO-AFI to FMM state. Similar behavior is observed for STO sample where resistivity is measurable at zero field and shows sharp changes in resistivity above 1.5 Tesla magnetic field indicating a field induced metal insulator transition. Besides 4 orders of magnitude change in virgin curve, resistivity during field reducing cycle from 8 Tesla to 0 Tesla remains almost constant, suggesting absence of reverse transformation to CO-AFI state. In fact there is almost negligible variation in resistivity for the entire envelope curve. This behavior, open hysteresis loop and virgin curve lying outside envelope curve in resistivity are similar to that observed in $Pr_{0.5}Ca_{0.5}Mn_{0.975}Al_{0.025}O_3$ (PCMAO) and some of the intermetallics [16, 20, 21]. There it is attributed to hindered kinetics of first order transition at low temperature. In our case we obtain almost AFI state when 5 K is reached in zero field condition whereas we obtain almost FM state at 5K after field cycling.

To address which of these state (CO-AFI or FMM) are equilibrium state, we used CHUF (cooling and heating in unequal field) protocol as has been used earlier to address such issues [20] and is shown in figure 4 and 5. Under this protocol in one case (figure 4) the sample is cooled in different fields that is lower or higher than the warming field and then warmed in one field (2 Tesla) while in the second case (figure 5) the sample is cooled in one field and then warmed in different field values lower and higher then the cooling field value. As shown in figure 4, the ρ(T) curve in warming in presence of 2T and cooled in a field of 1 T showed a flat region for all the three samples at lowest temperature followed by a sharp decrease indicating CO-AFI to FMM transformation. The ρ(T) than starts increasing again finally showing FMM to CO-AFI transition around 150 K. Whereas, for cooling fields of 2T and 4 T resistivity curves shows only one transition around 150 K. As mentioned before when cooled in zero field to 5K the system behaves as phase separated states, a combination of AFI and FMM phases and cooling in different fields changes the phase fraction of the two phases as reflected in the different values of the resistivity at 5 K. The main feature that emerges from this figure is below ~30 K. When the samples are warmed in fields that are lower than the cooling fields, the resistivity shows an anomalous flat region followed by a sharp decrease before increasing with increasing temperature. This is analogous to the devitrification process observed in classical glass transition. This behavior indicates presence of a blocked CO-AFI state. This blocked glassy state melts on increasing temperature and crystallizes in the stable FMM state resulting in sharp fall in resistivity. On the other hand ρ vs T curves under magnetic fields that are equal or higher than the cooling fields show behavior typical of a FMM phase. We can thus conclude that FMM state is the equilibrium stable state of this system while CO-AFI state is the metastable glassy state that coexists with it when measurements are performed under small fields. It also shows that one can tune glass-like CO-AFI phase fraction at 5 K by varying cooling field which is analogues to pressure in liquid to glass transition. A closer look at the devitrification curves [Figure 4 (cooled in 1 T, warmed in 2 T)] for all the three films shows that the devitrification process is completed at lower temperature in film on STO, as compared to films on LAO and NGO. This lower value of $T_g$ results in the part transformation of CO-AFI phase to FMM phase that is observed in zero field measurements for films on STO (see fig 2(a)).

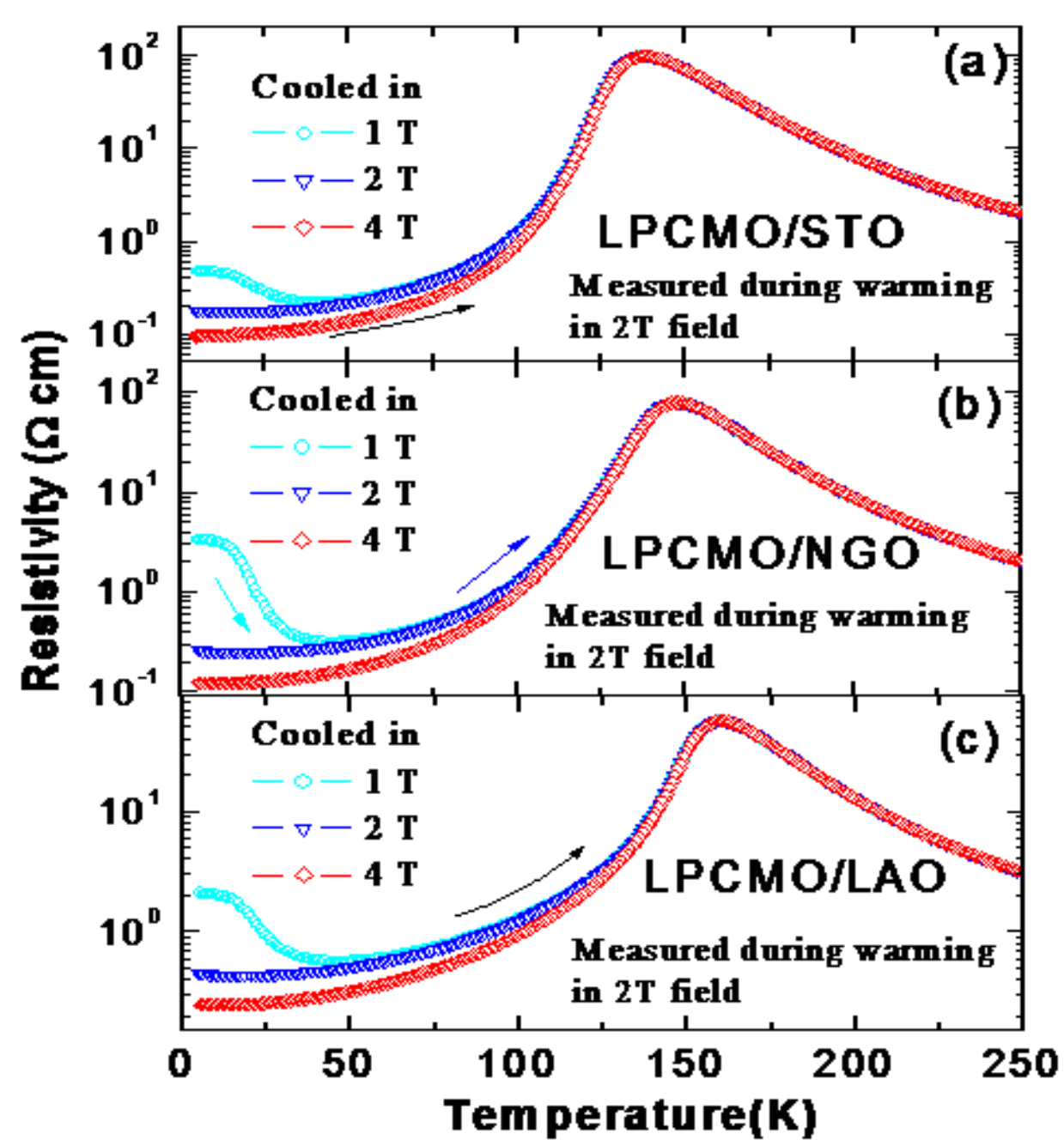

**Figure 4 :** Resistivity as function of temperature during warming in the presence of 2 T magnetic field for (a) LPCMO/STO, (b) LPCMO/NGO and (c) LPCMO/LAO. For each of these measurements samples are cooled in different magnetic field of 1 T, 2 T and 4 T respectively.

In liquid to glass transition, besides cooling or quenching rate, pressure has been used as an additional parameter to form glasses, like vitrification of Ge with pressure [29]. Similar to these studies we used magnetic field to tune glass like AF-I states. The AF-I phase obtained as a dominant state on cooling in H=0, shows all the characteristics of a glassy state including devitrification on warming. When the samples are cooled in constant field and warmed in different fields, the devitrification of the blocked glassy state is highlighted and is shown in figure 5. Cooling the sample from 300 K to 5 K in a field of 0.5 Tesla results in coexisting phases; a stable equilibrium phase and metastable glass state. Now warming the sample in different fields provides an opportunity of studying the process of devitrification behavior of the magnetic glass and dependence of $T_g$ on H. It is observed that when warmed in fields lower than or equal to the cooling field devitrification process is not observed and the heating curve shows monotonous increase in resistivity with increasing temperature. On the other hand when the sample is warmed in fields higher than cooling field, devitrification of glass-like CO-AFI phase to crystalline FMM phase is seen through sharp decrease in ρ that starts at 17 K and terminates at ~ 46 K, with the mid point of ρ ($T_g$ at 1 Tesla) being at ~ 25.6 K. The re-entrant transition, corresponding to melting of FMM phase, is seen by a sharp increase in ρ, with the mid-point of transition (T** at 1 Tesla) being at 114 K. The sample is again cooled from 300 K to 5K in a cooling field of $H_C$ = 0.5 Tesla to obtain the same initial state and CO-AFI phase fraction and then warmed from 5 K after increasing field to $H_W$=1.5 Tesla. The devitrification for FMM phase is again seen through a sharp drop in ρ that now starts at 11 K and terminates at about 41 K, with the mid point of ρ ($T_g$ at 1.5 Tesla) being about 18.5 K. The mid-point of the re-entrant transition, corresponding to melting (T** at 1.5 Tesla) is 121 K. The same procedure is repeated but now with warming field of 2 Tesla and resulted in drop in ρ starting at 6 K itself and terminates at 39 K, with the mid point of ρ ($T_g$ at 2 Tesla) being about 13.5 K. The mid-point of the re-entrant transition, corresponding to melting (T** at 2 Tesla) is 127 K. We see that T** rises with increasing field so that an isothermal increase of H would, in a specific range of T, convert the CO-AFI state to FMM. The value of dT**/dH must be consistent with Clausius Clapeyron relation since this is a first order transition. It is also noted that $T_g$ at 1 Tesla is 25 K, while that at 2 Tesla is 10 K. $T_g$ thus falls with increasing field, and an isothermal increase of H would, in a certain range of T, converts the glass-like CO-AFI phase to equilibrium FMM phase by devitrification. This is consistent with the qualitative condition imposed by Le Chatelier's Principle, in that increasing H takes the system to a state with higher M. Such a behavior of $T_g$ has also been observed for PCMAO [21].

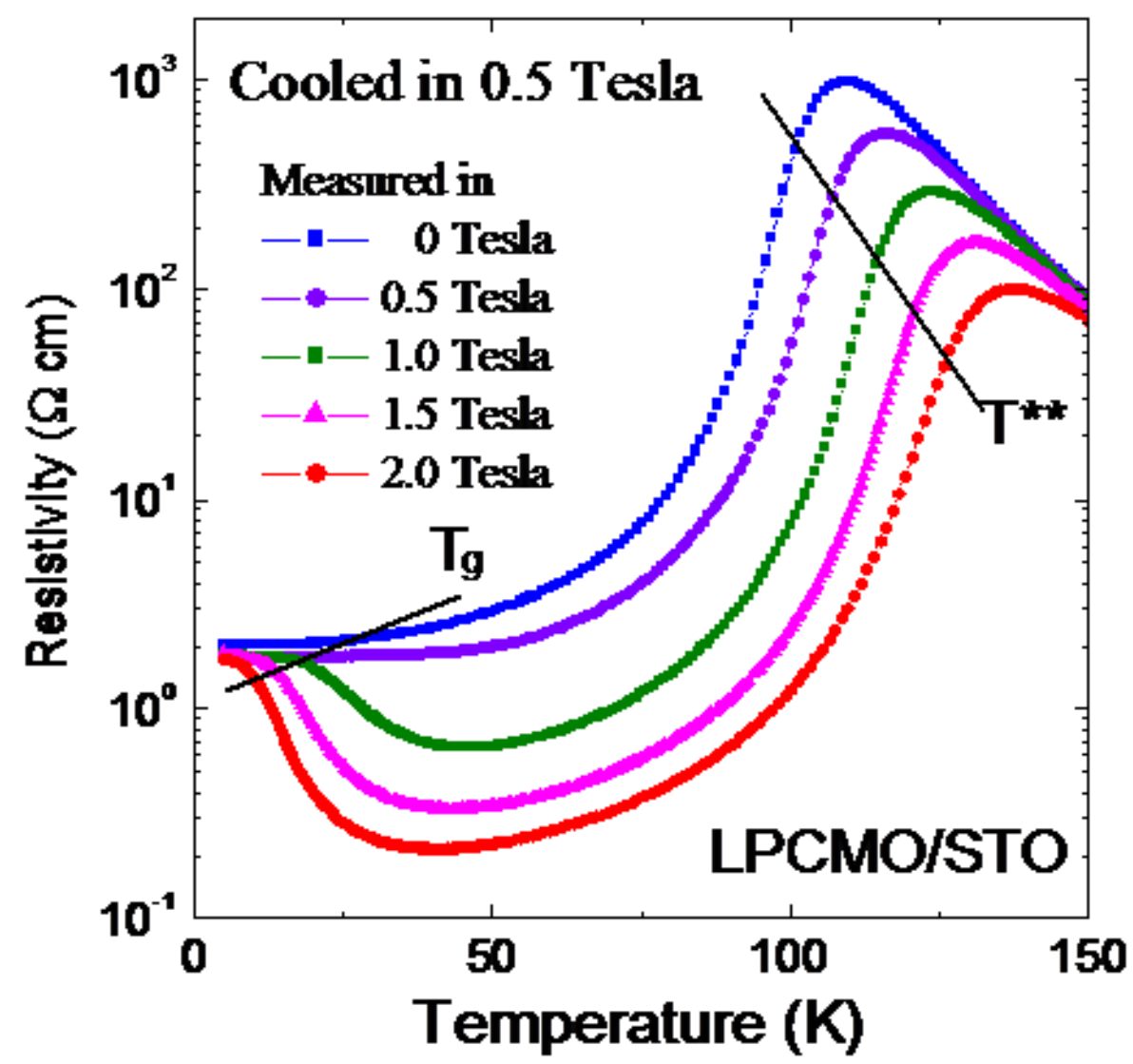

**Figure 5:** Resistivity as a function of temperature during warming in the presence of different magnetic field for LPCMO film deposited on STO. For each of these measurements sample is cooled in the presence of 0.5 Tesla magnetic field to 5 K and magnetic field changed to labeled field value isothermally. Devitrification at low temperature and reverse transformation to AFM state is highlighted. Dashed line shows Tg and T** line for these curve. See text for details.

In a classical liquid to glass transition, avoiding first order transition, cooling rate is very important. The faster the rate of cooling the more will be the fraction of glass phase in the total system. In fact we can totally avoid the liquid to crystal like first order transition with sufficiently high cooling rate that varies from system to system. In order to elucidate the effect of cooling rate on the present magnetic glass like phase we carried out cooling rate dependent ρ vs T measurements on LPCMO/STO sample that is shown in figure 6. The upper curve shows warming ρ(T) curve after cooling from 300 K to 5 K with a rate of 1.2 K/minute. The next curve represents ρ(T) cooling curve measured with a rate of 0.5 K/minute. It is observed that the value of ρ at 5 K decreases as the rate of cooling is decreased indicating decrease in glass-like CO-AFI phase as the rate of cooling is decreased. In short the relative volume fraction of CO-AFI with respect to the FMM phase decreases when the system is cooled more slowly across the first order transition and the associated time scale is in hours. This observation corroborates well with the normal glass transition.

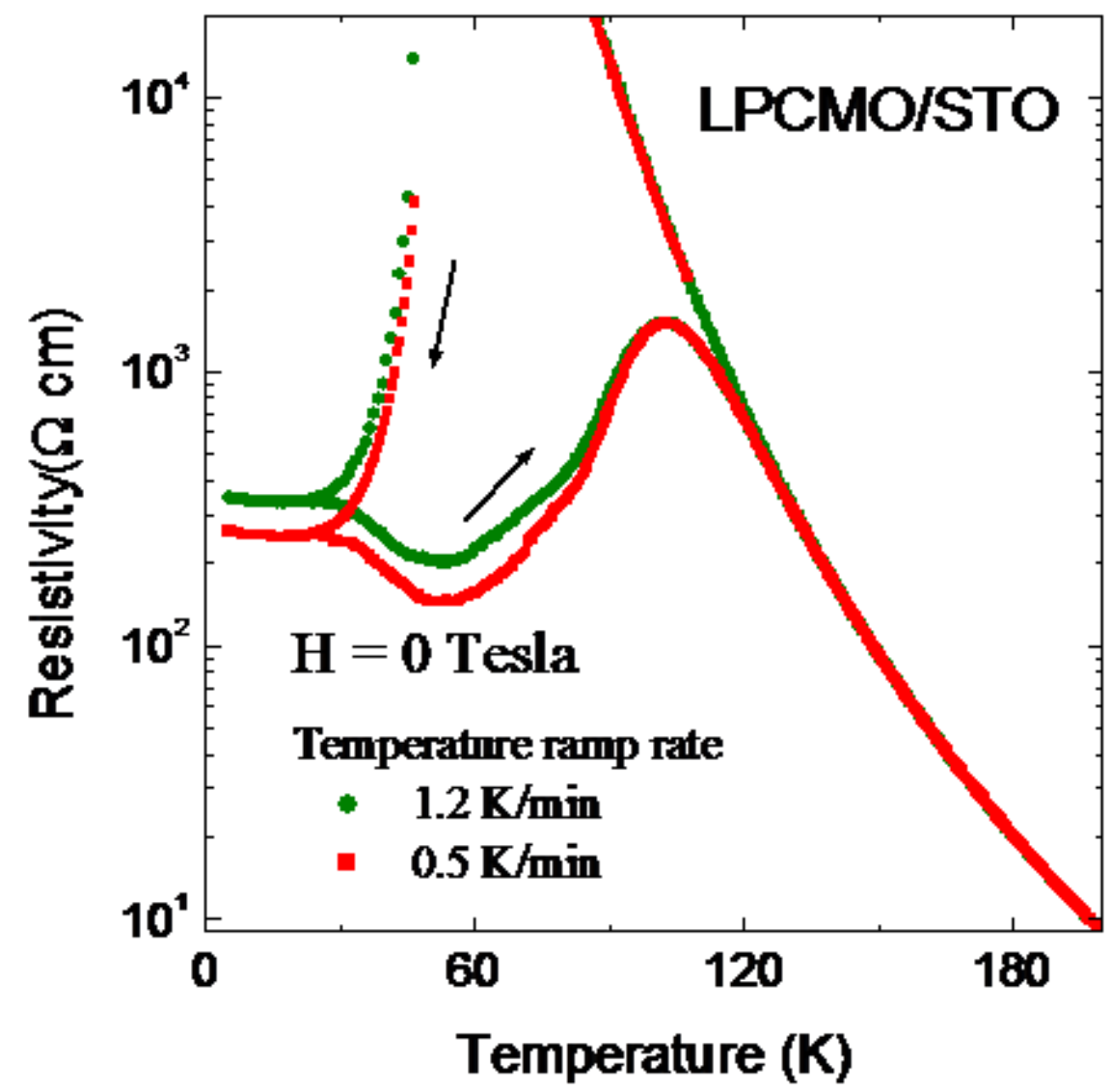

**Figure 6:** Resistivity as a function of temperature for $La_{5/8-y}Pr_yCa_{3/8}MnO_3$ (y=0.45) films deposited on STO with two different cooling rates. The resistivity at low T becomes smaller when the specimen is cooled slowly.

**Final Remarks:**

It is worth noting here that in all the films of LPCMO, deposited on different substrates like NGO, STO and LAO, the glass like arrest of kinetics is observed irrespective of strain. As mentioned before the film deposited on NGO is the least strained as the lattice parameter of the NGO (110) substrate (3.859 Å) nearly matches with the lattice parameters of the LPCMO bulk compound. On the other hand film deposited on STO shows an in-plain tensile strain while film on LAO shows in-plain compressive strain. The ρ(T) data for all the films show glass-like behavior below ~25 K. The substrate induced strain affects the magnitude of the resistivity in the films and there by affects the first order CO-AFI to FMM transition temperature. The long range CO state is enhanced with in-plane compressive strain and suppressed with in-plane tensile strain. Our observation is in contradiction to earlier studies carried out on y=0.3 samples where tensile strain enhances the CO-AFI state [30]. We believe that this may be due to difference in compositional variation; y=0.45 and y=0.3 compositions studied in the two cases. In y=0.45 composition the CO-AFI phase dominates and the system remains insulating down to lowest temperature while for y=0.3 composition FMM phase is in prominence at low temperature thus showing a insulator to metal transition at depleted temperatures [22]. However, the first order metal-insulator transition temperature values and hysteresis in heating and cooling under magnetic field of 2 T are very close for the three films (see figure 2). This clearly shows that the 'kinetic arrest' across the first order transition retains its importance in shaping the large scale phase coexistence and at least in LPCMO system seems to win over external effects like strain.

## Conclusions:

In summary, we performed an investigation of the low temperature transport properties of a micrometer length scale dynamic phase separated system, $La_{5/8-y}Pr_yCa_{3/8}MnO_3$ (y=0.45) thin films grown on NGO, STO and LAO substrates with emphasis on understanding the roles of strain and 'kinetic arrest' across first order transition. We have shown that phase separation is sensitive to the influence of the substrate induced strain. The long range CO-AFI state gets strength with compressive in-plane strain and suppressed with tensile strain. The resistivity vs. temperature measurements under cooling and heating in unequal fields (CHUF) protocol showed that the kinetics of first order phase transition is arrested across CO-AFI to FMM transition and the arrested state behaves as a magnetic glass. Like structural glasses, these magnetic glass-like phase shows evidence of devitrification of the arrested CO-AFI phase to equilibrium FMM phase with isothermal increase of magnetic field and/or iso-field warming. Using the CHUF protocol it is shown that CO-AFI is arrested metastable state while FMM is the equilibrium state of the system. The cooling rate dependent resistivity measurements further supported the existence of glass-like state in the system. Such a glass-like state and large scale dynamic phase separation is observed in all the three film irrespective of quite a different substrate induced strain. Thus in LPCMO 'kinetic arrest' dominates over strain in shaping large scale phase separation.

**Acknowledgements:** We acknowledge Suresh Bhardwaj for help in x-ray diffraction measurements and Sachin Kumar for help in resistivity measurements.

## References

[1] E. Dagotto New J. of Phys. **7,** 67 (2005).
[2] E Dagotto, Science **309**, 257 (2005).
[3] Liuwan Zhang, Casey Israel, Amlan Biswas, R. L. Greene, and Alex de Lozanne Science **298,** 805 (2002).
[4] M. Uehara and S.-W. Cheong Europhys. Lett. **52,** 674 (2000).
[5] P.A. Sharma, Sung Baek, Kim, T.Y. Koo, S Guha and S-W. Cheong Phys. Rev. B **71**, 224416 (2005).
[6] Tara Dhakal, Jacob Tosado and Amlan Biswas Rhys. Rev. B **75**, 092404 (2007).
[7] S. Chaudhuri and R.C. Budhani, Euro Phys. Lett. **81,** 17002 (2008).
[8] L. Ghivelder and F. Parisi Phys. Rev. B **71,** 184425 (2005).
[9] J. Sacanell et al Phys. Rev. B **73**, 014403 (2006); F. Macia et al Phys. Rev. B 76, 174424 (2007); F. Macia et al Phys. Rev. B **77**, 012403 (2008).
[10] K.H. Ahn, T. Lookman and A.R. Bishop Nature **428**, 401 (2004).
[11] Vijay B. Shenoy, Tribikram Gupta, H.R. Krishnamurthy, and T.V. Ramakrishnan Phys. Rev. B **80**, 125121 (2009); Vijay B. Shenoy, Tribikram Gupta, H.R. Krishnamurthy, and T.V. Ramakrishnan Phys. Rev. Lett. **98**, 097201 (2007).
[12] M.K. Chattopadhyay, S.B. Roy and P. Chaddah, Phys. Rev. B **72**, 180401R (2005), and references therein.
[13] A. Banerjee, K. Mukherjee, Kranti Kumar and P. Chaddah, Phys. Rev. B **74**, 224445 (2006); A. Banerjee, A K Pramanik, Kranti Kumar and P Chaddah, J. Phys.: Condens. Matter **18**, L605 (2006).
[14] W. Wu, C. Israel, N. Hur, P. Soonyong, S.-W. Cheong and A. De Lozane, Nat. Mater. **5**, 881 (2006).
[15] F. Macia, G. Abril, A. Hernndez-Mnguez, J.M. Hernandez, J. Tejada, and F. Parisi, Phys. Rev. B **76**, 174424 (2007).
[16] S.B. Roy, M.K. Chattopadhyay, P. Chaddah, J.D. Moore, G.K. Perkins, L. F. Cohen, K. A. Gschneidner and V.K. Pecharsky, Phys. Rev. B **74**, 012403 (2006); S.B. Roy, M.K. Chattopadhyay, A Banerjee, P. Chaddah, J.D. Moore, G.K. Perkins, L. F. Cohen, K. A. Gschneidner and V.K. Pecharsky, Phys. Rev. B **75**, 184410 (2007)
[17] P. Chaddah, Kranti Kumar and A. Banerjee, Phys. Rev. B **77**, 100402R (2008)
[18] Pallavi Kushwaha, R. Rawat and P. Chaddah J. Phys.: Condens. Matter **20**, 022204 (2008); Pallavi Kushwaha, Archana Lakhani, R. Rawat A. Banerjee and P. Chaddah Phys. Rev. B **79**, 132402 (2009).
[19] A. Banerjee, R. Rawat, K. Mukherjee, and P. Chaddah, Phys. Rev. B **79**, 212403 (2009)
[20] A Banerjee, Kranti Kumar and P Chaddah, J. Phys.: Condens. Matter **21**, 026002 (2009). S. Dash, A. Banerjee and P. Chaddah, solid state commun. **148**, 336 (2008).
[21] Archana Lakhani, Pallavi Kushwaha, R. Rawat, Kranti Kumar, A. Banerjee and P. Chaddah, J. Phys.: Condens. Matter **22**, 032101 (2010)).
[22] M. Uehara, S. Mori, C.H. Chen and S.-W. Cheong, Nature (London) **399**, 560 (1999).
[23] P.A. Sharma, S. El-Khatib, I. Mihut, J.B. Betts, A. Migliori, S. B. Kim, S. Guha, and S.-W. Cheong Phys. Rev. B **78**, 134205 (2008).
[24] Kranti Kumar, A.K. Pramanik, A. Banerjee, P. Chaddah, S.B. Roy, S. Park, C.L. Zhang, and S.-W. Cheong **73**, 184435 (2006).
[25] A. Yakubovskii, K. Kumagai, Y. Furukawa, N. Babushkina, A. Taldenkov, A. Kaul, and O. Gorbenko, Phys. Rev. B **62**, 5337 (2000).
[26] A. Garashenko, Y. Furukawa, K. Kumagai, S. Verkhovskii, K. Mikhalev, and A. Yakubovskii, Phys. Rev. B **67**, 184410 (2003).

[27] H. J. Lee, K.H. Kim, M. W. Kim, T. W. Noh, B. G. Kim, T. Y. Koo, S.-W. Cheong, Y. J. Wang, and X. Wei, Phys. Rev. B 65, 115118 (2002).
[28] K.H. Kim, M. Uhera, C. Hess, P.A. Sharma, and S.-W. Cheong, Phys. Rev. Lett. 84, 2961 (2000).
[29] M.H. Bhat, V. Molinero, E. Soignard, V.C. Solomon, S. Sastry, J.L. Yarger and C.A. Angell, Nature **448**, 787 (2007).
[30] Dane Gillaspie, J.X. Ma, Hong-Ying Zhai, T.Z. Ward, Hans M. Christen, E.W. Plummer and J. Shen J. Appl. Phys. **99**, 08S901 (2006).